\documentclass[conference]{IEEEtran}

\usepackage{xcolor}
\usepackage{comment}
\usepackage{graphicx}
\usepackage{multirow}
\usepackage{caption}
\usepackage{subcaption}
\ifCLASSINFOpdf
\else
\fi
\begin{document}
%
\title{\LARGE \bf White Mirror: Leaking Sensitive Information from Interactive Netflix Movies using Encrypted Traffic Analysis\vspace{-10pt}}

\author{\IEEEauthorblockN{Gargi Mitra, Prasanna Karthik Vairam, Patanjali SLPSK, Nitin Chandrachoodan, Kamakoti V}
\IEEEauthorblockA{Indian Institute of Technology Madras\\
Email: \{gargim@cse,pkarthik@cse,slpskp@cse,nitin@ee,kama@cse\}.iitm.ac.in\vspace{-10pt}}}


%


\maketitle

\begin{abstract}
Privacy leaks from Netflix videos/movies is well researched. Current state-of-the-art works have been able to obtain coarse-grained information such as the genre and the title of videos by passive observation of encrypted traffic. However, leakage of fine-grained information from encrypted traffic has not been studied so far. Such information can be used to build behavioral profiles of viewers.

On 28th December 2018, Netflix released the first mainstream interactive movie called `Black Mirror: Bandersnatch'. In this work, we use this movie as a case-study to show for the first time that fine-grained information (i.e., choices made by users) can be revealed from encrypted traffic. We use the state information exchanged between the viewer's browser and Netflix as the side-channel. To evaluate our proposed technique, we built the first interactive video traffic dataset of $100$ viewers; which we will be releasing. Preliminary results indicate that the choices made by a user can be revealed $96\%$ of the time in the worst case.
\end{abstract}

\IEEEpeerreviewmaketitle

\section{Introduction}
Recently, Netflix released the first mainstream interactive movie called {\em Black Mirror: Bandersnatch}. In this movie, viewers are allowed to create their own story-line by choosing one of the on-screen options presented to them. For instance, the viewers are asked choice-questions such as `Frosties or sugar-puffs?', `visit therapist or follow Colin?', `throw tea over computer or shout at dad?'. Depending on the viewer's choice, the corresponding segment of the movie gets played. Interestingly, the choices made and the path followed can potentially reveal viewer information that ranges from benign (e.g., their food and music preferences) to sensitive (e.g., their affinity to violence and political inclination). Although this information is available to Netflix, they are bound by legal clauses that prevent them from misusing it. Today, to prevent this information from leaking to unauthorized parties, Netflix uses end-to-end encryption. Recent advancements in the domain of encrypted network traffic analysis make it possible to infer basic information about the preferences of Netflix viewers. For instance, prior works~\cite{reed2017identifying, schuster2017beauty} were able to infer the title of videos watched by viewers and their preferred genre. 
In this work, we raise the following research question: `Do interactive movies leak sensitive fine-grained information about the viewers to passive eavesdroppers due to the choices that they make while watching the movie?'.

\noindent \textbf{Contributions. (1)} We present the first traffic analysis technique for interactive videos that can leak more information than non-interactive videos. \textbf{(2)} We present the first dataset for encrypted traffic analysis of interactive videos.

\section{Related Work}
The objectives of existing research works~\cite{reed2017identifying,schuster2017beauty,li2018silhouette} have so far been limited to discerning video content from other traffic, identifying the titles of videos and the genre preferred by a viewer from encrypted traffic of {\em conventional} videos. However, the objective of our work is to see if more fine-grained private information about a viewer can be leaked from encrypted interactive video traffic.

Existing techniques from the literature are not suitable for encrypted interactive video traffic analysis due to the following fundamental reason: inter-video features cannot be used to differentiate between segments from the same video. For instance, prior works~\cite{reed2017identifying} have used bitrate as a feature to differentiate between two video streams. However, in the context of an interactive video, the bitrate of chunks pertaining to each choice will be the same and hence cannot be used to distinguish between two video segments. In this work, we identify an intra-video side-channel and show that it holds across various operating conditions.

\section{Proposed Traffic Analysis Technique}
\begin{figure}
    \centering
    \includegraphics[width=0.5\textwidth]{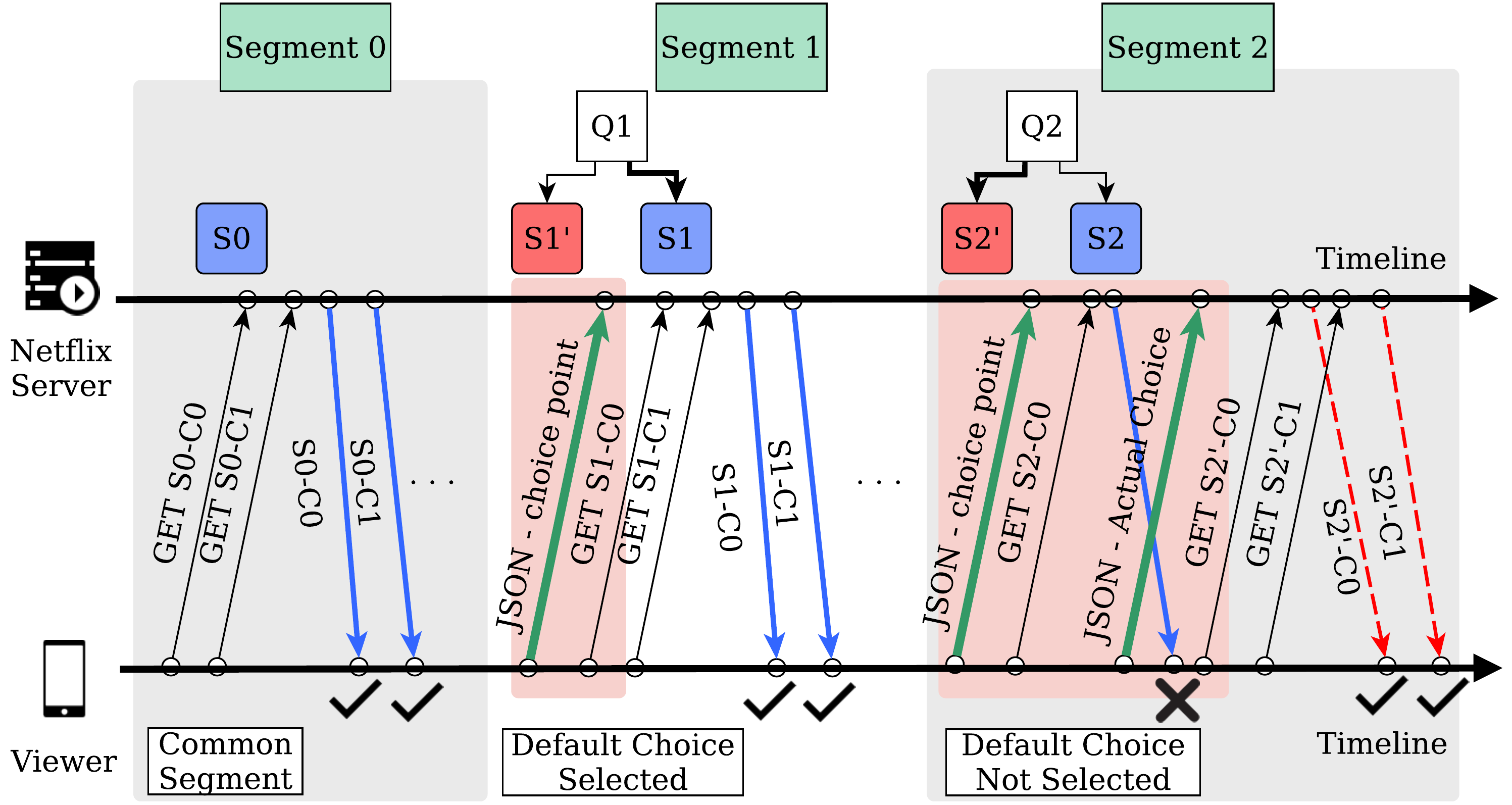}
    \caption{\small The streaming process of Black Mirror: Bandersnatch\vspace{-15pt}}
    \label{fig:bandersnatch}
\end{figure}

In this section, we first describe the streaming process of Bandersnatch. Due to the non-linearity of the script, the streaming process is check-pointed at each choice-question in the movie unlike conventional Netflix videos that stream continuously. The content of Bandersnatch is divided into several segments, each corresponding to one path segment in the script, with each segment consisting of multiple chunks. Figure~\ref{fig:bandersnatch} shows an example where a viewer makes the first two choices. The first segment of the movie (i.e., Segment 0) is common for all viewers, and its chunk streaming ends when the viewer is presented with the first choice-question $Q1$. At this point, the viewer's browser sends a JSON file (type-1 JSON file) to the server indicating that the viewer has encountered $Q1$. The viewers are then given ten seconds to choose one out of two options. 
Our experiments revealed that Netflix considers one of the choices to be the default branch and prefetches chunks belonging to the default segment. We denote the default choice for question $Qi$ as $Si$ and the non-default choice as $Si'$. If the viewer chooses $Si$, then the streaming continues uninterrupted. However, if the choice $Si'$ is chosen, the prefetching for $Si$ stops and a request for $Si'$ is sent. 

\begin{table}[!t]
\centering
\begin{tabular}{|p {1.2cm}|c|p{3cm}|}
\hline
                             \textbf{Conditions}  & \textbf{Attribute}  & \textbf{Value}                  \\ \hline
\multirow{5}{*}{Operational}     & Operating System    & Windows, Linux, Mac             \\ \cline{2-3} 
                               & Platform            & Desktop, Laptop                 \\ \cline{2-3} 
                               & Traffic Conditions  & Morning, Noon and Night         \\ \cline{2-3} 
                               & Connection Type     & Wired, Wireless                 \\ \cline{2-3} 
                               & Browser             & Google-chrome, Firefox          \\ \hline
\multirow{4}{*}{Behavioral} & Age-group           & $<$ 20, 20-25,25-30, $>$ 30 \\ \cline{2-3} 
                               & Gender              & Male,Female,Undisclosed       \\ \cline{2-3} 
                               & Political Alignment &     Liberal, Centrist, Communist, Undisclosed                            \\ \cline{2-3} 
                               & State of Mind       &    Happy, Stressed, Sad, Undisclosed                             \\ \hline
\end{tabular}
\caption{Attributes of the IITM-Bandersnatch Dataset\vspace{-15pt}}
\label{tab:dataset}
\end{table}
In the example from Figure~\ref{fig:bandersnatch}, the viewer selects the default choice $S1$ for $Q1$. The streaming continues uninterrupted until $Q2$ appears on-screen. Like before, a type-1 JSON file is sent from the viewer's browser to Netflix at this point. The viewer selects the non-default option $S2'$ for Q2. In this case, another JSON file (i.e., of type-2) will be sent. Although some chunks corresponding to $S2$ are pre-fetched, they will be discarded and the chunks corresponding to $S2'$ will be streamed. Therefore, we can conclude that the number and type of JSON files sent indicate the choice made by the viewer. However, identifying the two types of JSON files from encrypted network traffic is challenging.

Our experiments revealed that the packets carrying the encrypted type-1 and type-2 JSON files can be distinguished from other packets by their SSL record lengths which are visible even from encrypted traffic. This observation was found to be consistent across various operating systems, browsers, devices, connection media, and network conditions. Hence, we use {\em `SSL record lengths of client packets'} as a side-channel to infer the choices made by the viewer while watching Bandersnatch.

\section{The Interactive Video Traffic Dataset}

In this section, we describe our dataset, called IITM-Bandersnatch, that we use to evaluate the proposed traffic analysis technique. The dataset comprises of data points of the form \{encrypted traces, ground truth choices\}. To collect each data point, we asked the viewer to watch Bandersnatch from the beginning and note down the choices made by them. At the same time, we collected the encrypted network traffic. As of now, our dataset contains information corresponding to $100$ viewers who volunteered for this study.

The robustness of a traffic analysis technique can only be evaluated if the dataset has a diverse set of representative data points. In this dataset, we account for diversity by considering various operating conditions such as multiple Operating Systems, devices, browsers, connection type, and network conditions. Table~\ref{tab:dataset} summarizes the operating conditions considered.
\begin{figure}[!t]
\begin{minipage}{0.24\textwidth}
    \includegraphics[scale=0.3]{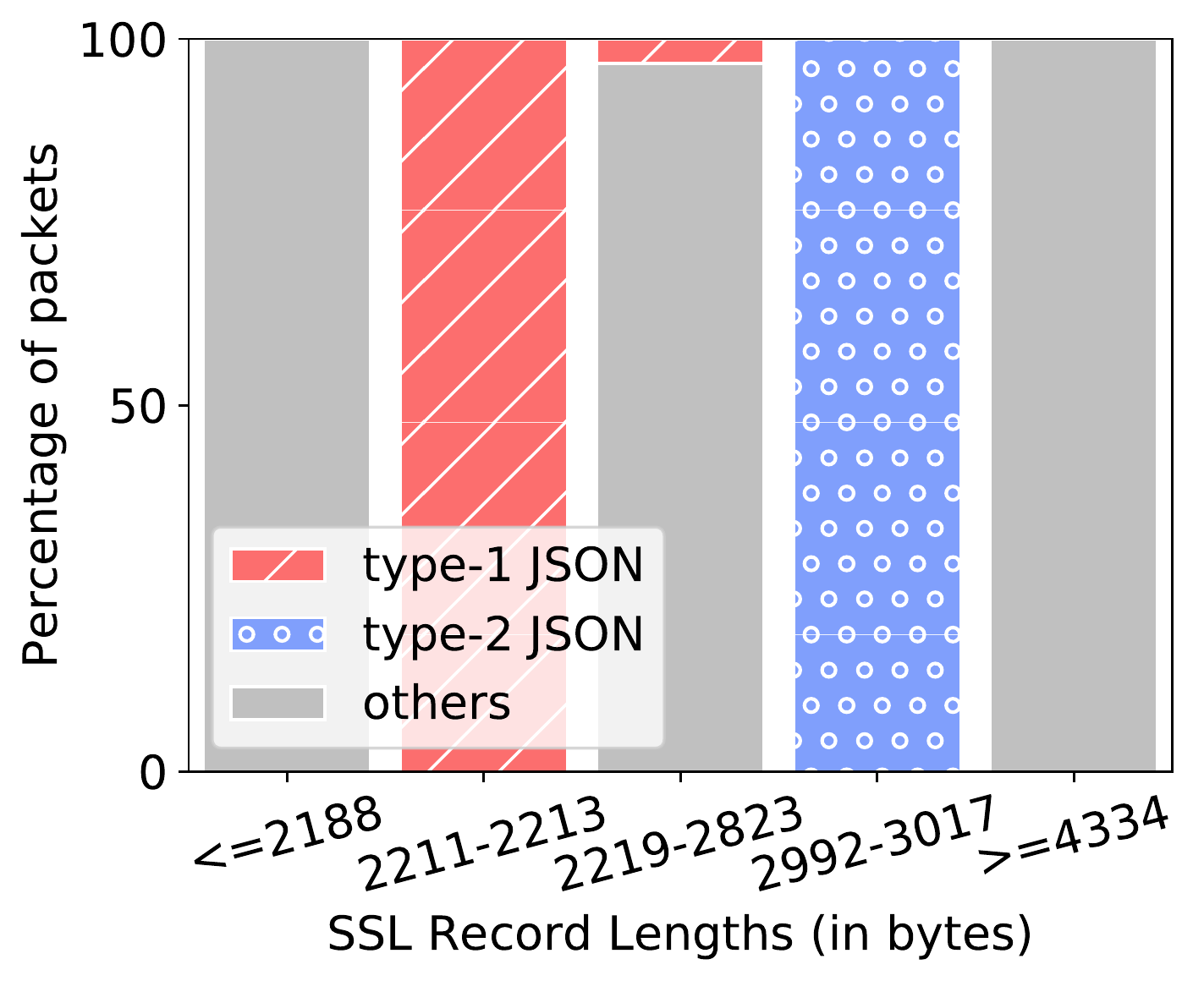}\vspace{-15pt}
    \label{fig:dfeb}
\end{minipage}
\begin{minipage}{0.24\textwidth}
    \includegraphics[scale=0.3]{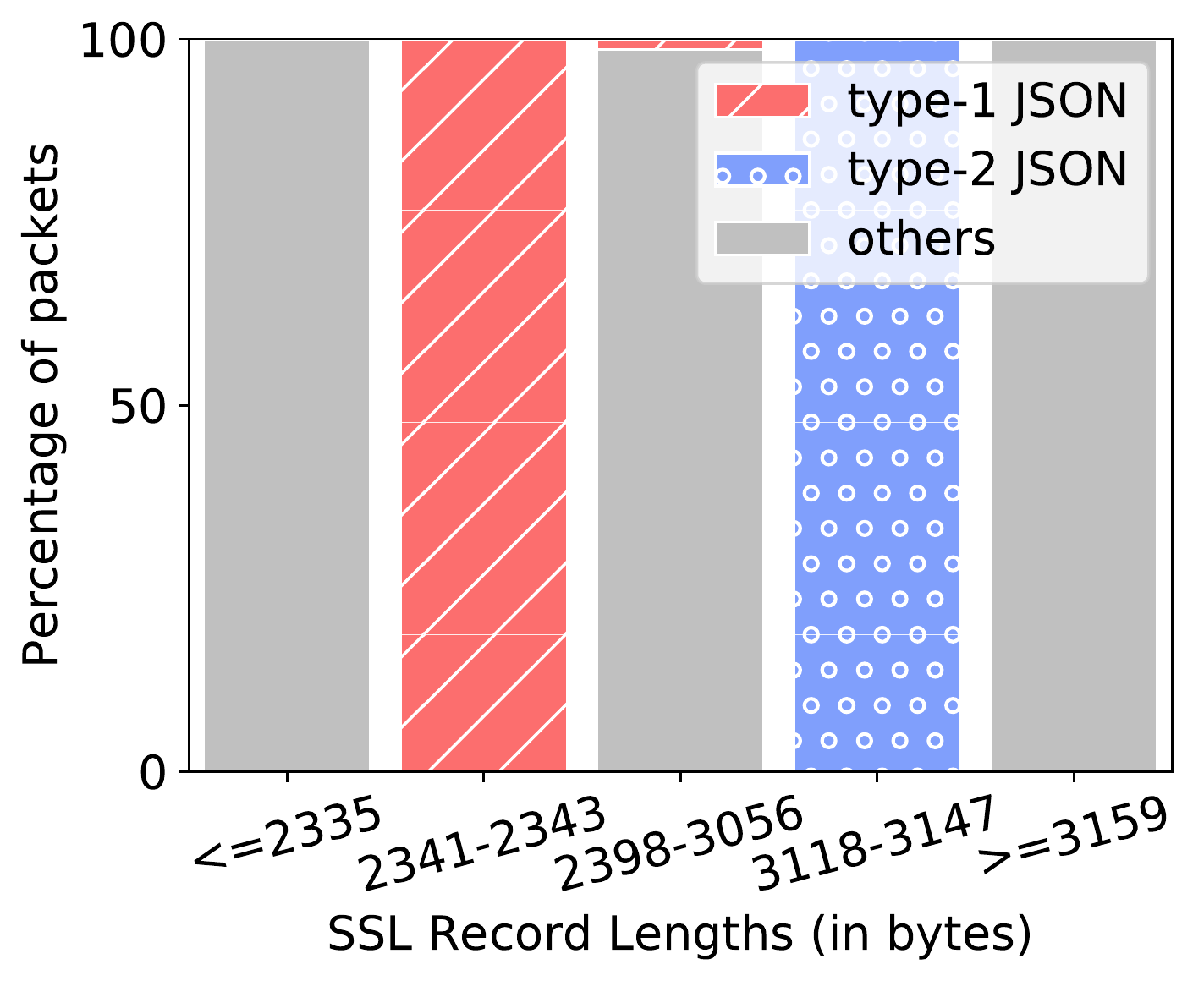}\vspace{-15pt}
    \label{fig:dfew}
\end{minipage}
\caption{\small SSL record length distribution for (Desktop, Firefox, Ethernet, Ubuntu) and (Desktop, Firefox, Ethernet, Windows)\vspace{-15pt}}
\label{fig:results}
\end{figure}

To enable behavioral sciences researchers to evaluate their techniques, we have also collected the behavioral state of the viewers who participated in the study. 

\section{Results} 
We conducted our preliminary experiments on the encrypted traffic captured during $10$ different viewing sessions of the movie 'Black Mirror: Bandersnatch'. In each case, the movie was viewed by different people under different combinations of operational and network conditions. Figure~\ref{fig:results} shows the distinguishability of packets carrying type-1 and type-2 JSON files using the SSL record lengths of client-side application packets obtained from the encrypted traffic of viewers under two different operational conditions. This helped us to identify the two types of JSON files with $96\%$ accuracy and hence the choices made by the viewers. 
\section{Conclusions and Future Directions}
In this work, we present a novel side-channel attack on interactive videos. We show that the identified side-channel holds for various operational and behavioral conditions. We also present the IITM-Bandersnatch dataset which contains the behavioural and operational data of 100 viewers.

\noindent \textbf{Countermeasures.} An easy fix for the problem would be to either split the JSON file or to compress it so that it becomes indistinguishable. However, there could be timing side-channels that may still exist even after this fix.

\noindent \textbf{Extending the Dataset.} The set of behavioral attributes in our dataset are limited. Extending the dataset with information pertaining to more relevant behavioral traits is an interesting direction.

\noindent \textbf{High-level Implications.} Our work is a modest attempt at identifying the viewer choices from encrypted traffic. We reach out to the research community to use this information for behavioral studies.


\bibliographystyle{IEEEtran}

\end{document}